\documentclass[conference]{IEEEtran}
\usepackage{amsmath}
\usepackage{amssymb}
\usepackage{graphicx}
\begin{document}
\title{On Large Deviation Property of Recurrence Times}
\author{\IEEEauthorblockN{Siddharth Jain}
\IEEEauthorblockA{Department of Electrical Engineering\\
Indian Institute of Technology Kanpur\\
Kanpur, India 208016\\
Email: sidjain@iitk.ac.in}
\and
\IEEEauthorblockN{Rakesh K. Bansal}
\IEEEauthorblockA{Department of Electrical Engineering\\
Indian Institute of Technology Kanpur\\
Kanpur, India 208016\\
Email: rkb@iitk.ac.in}}
\maketitle
\begin{abstract}
We extend the study by Ornstein and Weiss on the asymptotic behavior of the normalized version of recurrence times and establish the large deviation property for a certain class of mixing processes. Further, an estimator for entropy based on recurrence times is proposed for which large deviation behavior is proved for stationary and ergodic sources satisfying similar mixing conditions.
\end{abstract}
\section{Introduction}
For a stationary and ergodic source with finite alphabet, the asymptotic relationship between probability of an $n$ length sequence and entropy has been well established by Shannon-McMillan-Breiman Theorem~\cite{Shields96}. Later, Ornstein and Weiss~\cite{Ornstein93} established a similar expression relating recurrence times to entropy. Kontoyiannis~\cite{Kontoyiannis98} related recurrence times and probability of an $n$ length sequence for Markov sources by showing that $\lim_{n\rightarrow \infty}\log[R_n(X)P(X_1^n)] = o(n^\beta)~a.s.$, for any $\beta > 0$. Here, $R_n(X)$ and $P(X_1^n)$ represent random variables for recurrence time and probability of an $n$-length block generated by a source $X$, respectively. Further, in \cite[Corollary 2]{Kontoyiannis98}, he also identified a class of processes for which central limit theorem (CLT) and law of iterated logarithm (LIL) hold true for recurrence times.

The question of large deviations for Shannon-Mcmillan-Breiman Theorem has been successfully answered in literature under certain mixing conditions~\cite{Shields96}. Motivated by Kontoyiannis' results and the satisfaction of large deviation property for Shannon-McMillan-Breiman Theorem, it is natural to ask under what conditions the asymptotic recurrence times relation satisfies large deviation property. Chazottes and Ugalde~\cite{Chazottes2005} in 2005 established partial large deviations results on recurrence times for Gibbsian sources. In this paper, we have found a class of processes for which large deviation property holds for recurrence times.

For an i.i.d source, Shannon-McMillan-Breiman Theorem satisfies large deviation property by direct application of Cramer's Theorem~\cite{Dembo}. However for Ornstein and Weiss recurrence times result, even for an i.i.d. source Cramer's Theorem is not applicable. This makes the analysis of large deviation property for recurrence times non-trivial even for the i.i.d case. Hence, in order to answer the question of large deviations for recurrence times one needs to look more closely into the recurrence time statistics.

Maurer~\cite{Maurer92} studied the behavior of recurrence time statistics under the assumption of non-overlapping recurrence blocks for i.i.d sources. Later, Abadi and Galves~\cite{AbadiG04} studied a similar non-overlapping scenario for $\psi$-mixing processes and established an exponential bound on the recurrence time distribution. Moreover, they also brought out the contrast between overlapping and non-overlapping case. In the context of overlapping $R_n(x)$, there are several references that show convergence in distribution of $R_n(x)P(x_1^n)$ to an exponentially distributed random variable for a certain class of stationary and ergodic processes~\cite{Abadi04}\cite{Abadi09}\cite{Collet99}\cite{Galves97}\cite{Hirata99}. Kim~\cite{Kim12} also studied the behavior of conditional distribution of $R_n(X)P(X_1^n)$ given the $n$-length block $X_1^n = x_1^n$ and established an exponential bound on its distribution for these two classes of sources i) $\psi$-mixing, ii) $\phi$-mixing with summable coefficients. In this paper, in order to do our analysis we have used this exponential bound on conditional distribution established by Kim~\cite{Kim12}.

The rest of the paper is organized as follows. In section II we state preliminary results on recurrence time statistics and mixing processes. In section III, we state our main theorems for the large deviation property of recurrence times. In section IV, we give proofs of these theorems and their corollaries. In section V, we define an estimator for entropy based on recurrence times and prove large deviation property for it. In section VI, we present our conclusion.

\section{Preliminaries}
Let $\{X_n\}_{n=-\infty}^{n = \infty}$ be a stationary and ergodic process defined on the space of infinite sequences $(\mathcal{A}_{-\infty}^{\infty}, \sigma, P)$. Here $\mathcal{A}$ is a finite set of alphabets, $\sigma$ is the sigma field generated by finite dimensional cylinders and $P$ is the probability measure.
For simplicity of notation, we will use $X$ for $\{X_n\}_{n = -\infty}^{n = \infty}$. \\
$X$ is called $\psi$-mixing if
\begin{equation}\label{psi}
\sup_{A\epsilon\sigma_{-\infty}^n,~ B\epsilon\sigma_{n+l}^\infty}\frac{|P(A\cap B)-P(A)P(B)|}{P(A)P(B)} \leq \psi(l).
\end{equation}
Here, $\psi(l)$ is a decreasing sequence converging to $0$ and $\sigma_i^j$ denotes the sigma algebra generated by $X_i^j = X_iX_{i+1}....X_j$ and it is called $\phi$-mixing if
\begin{equation}\label{phi}
\sup_{A\epsilon\sigma_{-\infty}^n,~ B\epsilon\sigma_{n+l}^\infty}\frac{|P(A\cap B)-P(A)P(B)|}{P(A)} \leq \phi(l).
\end{equation}
Here, $\phi(l)$ is a decreasing sequence converging to $0$. \\
Let $\{x_n\}_{n=-\infty}^{n=\infty}$ denote a particular realization of $X$. Now define, the first return time (recurrence time) of $x_1^n$ to be:
$$R_n(x) = \min\{j \geq 1:x_1^n = x_{-j+1}^{-j+n}\}.$$
As a dual of recurrence time $R_n(x)$, match length $L_m(x)$ is defined as follows:
$$ L_m(x) = \max\{j \geq  1:x_1^j = x_{-k+1}^{-k+j},~ k = 1, 2,...,m\}.$$
{\bf Observation 1}~\cite{Kontoyiannis98}{\bf:} $R_n(x) > m \Leftrightarrow L_m(x) < n. $\\
On asymptotic behavior of $R_n(x)$ and $L_m(x)$ following holds,
{\bf Ornstein and Weiss \cite{Ornstein93}}\\
For $X$ with entropy rate $H(X)$, with probability 1,
$$\lim_{n \rightarrow \infty}\frac{\log R_n(X)}{n} = H(X); ~~~~\lim_{m \rightarrow \infty}\frac{\log m}{L_m(X)} = H(X).$$
Further, in this paper unless stated otherwise, $i)$ we use $H$ to represent the entropy rate of the source $X$. $ii)$ A $\phi$-mixing process is assumed to be $\phi$-mixing in both forward and backward directions.\\
{\bf Kim's Theorem \cite{Kim12}}\\
For $X$ satisfying $\psi$-mixing condition or $\phi$-mixing condition with summable coefficients,
\begin{equation}\label{LB}
\begin{split}
P(R_n(X) > t|X_1^n = x_1^n) > \xi_x e^{-t\xi_xP(x_1^n)}(1-\\2\sqrt{C_x(\xi_xP(x_1^n)t\vee 1)})
~~~\forall~~ t > 0 .
\end{split}
\end{equation}
\begin{equation}\label{UB}
\begin{split}
P(R_n(X) > t|X_1^n = x_1^n) &< \xi_x e^{-t\xi_xP(x_1^n)}[1+ K(x,t)+\\&+ 2C_x(\xi_xtP(x_1^n)\vee 1)]~\forall~ t \geq \rho_x.
\end{split}
\end{equation}
where $ C_x = C\{\inf_{n\leq \triangle\leq 1/P(x_1^n)}[\triangle P(x_1^n)+ \ast(\triangle)]\}$ ($C>0$ is a constant, $\ast$ represents $\psi$ or $\phi$), $\rho_x = \frac{2\sqrt{C_x}}{(\sqrt{1+C_x}+\sqrt{C_x})\xi_xP(x_1^n)}$.$$ K(x,t) = 2\sqrt{C_x(t\xi_xP(x_1^n)\vee 1)(1+C_x(t\xi_xP(x_1^n)\vee 1))},$$   and $C_x \rightarrow 0 ~ (as ~ n \rightarrow \infty)$ and $\xi_x~\epsilon~[E_1, E_2], (0 < E_1<1<E_2<\infty).$ $a_1\vee a_2$ means $\max\{a_1,a_2\}$. Following additional properties as listed in \cite{Kim12} and originally proved in \cite{Collet99}\cite{Abadi01} hold for $\phi$-mixing processes:
\begin{enumerate}
\item For an exponentially $\phi$-mixing process,~ $\forall x_1^n ~\epsilon~ \mathcal{A}^n,$ there exists a positive constant $D_o$ and $\Gamma > 0$, s.t. $\forall n \geq n_o$
\begin{equation}\label{exp_phi}
C_x \leq D_oe^{-\Gamma n}.
\end{equation}
\item Let $B_n(s)$ be the set of $x_1^n ~\epsilon~\mathcal{A}^{n}$; such that $R_n(x) < \frac{n}{s}.$ Then, for any $\phi$-mixing process, there exists $s~\epsilon~\mathcal{N}$ ($\mathcal{N}$ being the set of natural numbers), and two positive constants $D_1$ and $d_1$ such that
\begin{equation}\label{4}
P(\{x: x_1^n ~\epsilon~B_n(s)\}) \leq D_1e^{-d_1n}.
\end{equation}
\item For exponentially $\phi$-mixing processes for every $x_1^n ~ \epsilon~ {\mathcal{A}}^n\backslash B_n(s)$
\begin{equation}\label{xi}
|\xi_x - 1| < D_2e^{-d_2n}.~~~~(for~n~large~enough)
\end{equation}
Here, $D_2$ and $d_2$ are constants.
\end{enumerate}
Now, we state a Lemma which is required in the proof of Theorem 4 stated in section III.
Let $A_1$, $A_2$ and $A_3$ be three sets such that $A_3 = A_1\cap A_2$. Suppose $P(A_1) > 1- p_1e^{-p_2n}$
and $P(A_2) > 1-q_1e^{-q_2n}$, where $p_1$, $p_2$, $q_1$ and $q_2$ are positive constants. Then, we have\\
{\bf Lemma 1:} $P(A_3) > 1- (p_1 + q_1)e^{-\min\{p_2,q_2\}n}$.\\ Lemma 1 is proved in the appendix.\\
{\bf Definition} \cite{Shields96} \\
$X$ is said to have exponential rates for entropy if for every $\epsilon~ > ~0$, we have
\begin{equation}\label{LDP_Sh}
P(\{x_1^n: 2^{-n(H+\epsilon)} \leq P(x_1^n) \leq 2^{-n(H-\epsilon)}\}) \geq 1- r(\epsilon,n).
\end{equation}
where $-\frac{1}{n}\ln r(\epsilon, n)$ is bounded away from 0 or in other words $r(\epsilon, n) = e^{-k(\epsilon)n}$, where $k(\epsilon)$ is a real valued positive function of $\epsilon$.\\
{\bf Theorem 1} \cite{Shields96}
\begin{enumerate}
\item I.I.D., ergodic Markov and $\psi$-mixing processes all have exponential rates for entropy.
\item An aperiodic and irreducible Markov chain is $\psi$-mixing.
\end{enumerate}
{\bf Remark 1:} In \cite{Shields96}, the $\psi$-mixing condition used is weaker as to what we have defined in Eq. (\ref{psi}). So, Theorem 1 also holds for processes satisfying the stronger $\psi$-mixing condition given in Eq. (\ref{psi}).\\
{\bf Theorem 2} \cite{Bradley}
\begin{enumerate}
 \item If a process is $\psi$-mixing, then it is also $\phi$-mixing.
 \item If a Markov Chain is $\phi$-mixing then it is exponentially $\phi$-mixing.
\end{enumerate}
{\bf Corollary 1:} From Theorem 1 and 2, it follows that
an aperiodic and irreducible Markov Chain is exponentially $\phi$-mixing and has exponential rates for entropy.\\
\section{Main Theorems}
{\bf Theorem 3:} For a process satisfying $\psi$-mixing condition or $\phi$-mixing condition with summable coefficients and with exponential rates for entropy,
$$P(\frac{\log R_n(X)}{n} > H+ \epsilon) \leq e^{-f(\epsilon)n} ~~~~  \forall~ n \geq N(\epsilon).$$
where, $f(\epsilon)$ is a real positive valued function for all $\epsilon~ > ~0 $ and $f(0) = 0$.\\
{\bf Corollary 2:}
Under the conditions of Theorem 3, we have
$$P(\frac{\log m}{L_m(X)} > H+ \epsilon) \leq e^{-f(\epsilon)\frac{\log m}{H+\epsilon}} ~~~~  \forall~ m \geq M(\epsilon).$$
{\bf Theorem 4:} For an exponentially $\phi$-mixing process,
$$P(\frac{\log R_n(X)}{n} < H-\epsilon) \leq e^{-g(\epsilon)n} ~~~~\forall~ n \geq N'(\epsilon).$$
where $g(\epsilon)$ is a real positive valued function for all $\epsilon~ > ~0 $ and $g(0) = 0$.\\
{\bf Corollary 3:} Under the conditions of Theorem 4, we have
$$P(\frac{\log m}{L_m(X)} < H- \epsilon) \leq e^{-g(\epsilon)\frac{\log m}{H-\epsilon}} ~~~~  \forall~ m \geq M'(\epsilon).$$
Theorem 3 and 4 are combined in the form of\\
{\bf Theorem 5 (Large Deviation Property for Recurrence Times)}\\
For an exponentially $\phi$-mixing process with exponential rates for entropy,
$$ P(|\frac{\log R_n(X)}{n} - H| > \epsilon) \leq 2e^{-I(\epsilon)n} ~~~~\forall n~ \geq N''(\epsilon). $$
where, $I(\epsilon)=\min\{f(\epsilon), g(\epsilon)\}$ and $N''(\epsilon)=\max\{N(\epsilon), N'(\epsilon)\}$.\\
{\bf Remark 2:} From, Corollary 1, it can be inferred that the quantity $\frac{\log R_n(X)}{n}$ for an aperiodic and irreducible Markov chain satisfies Large Deviation Property.
\section{Proofs}
\textbf{Proof of Theorem 3:}\\
Let $A_n^{(\delta)}$ be a set of n long sequences defined as,
$$ A_n^{(\delta)} = \{x_1^n: 2^{-n(H+\delta)} \leq P(x_1^n) \leq 2^{-n(H-\delta)}\}.$$Now,
\begin{equation*}
\begin{split}
&P(\frac{\log R_n(X)}{n} > H+ \epsilon) = P(R_n(X) > 2^{n(H+\epsilon)}) \\&= \sum_{y \epsilon \mathcal{A}^n}P(y)P(R_n(X) > 2^{n(H+\epsilon)}|X_1^n = y)\\
&= \sum_{y \epsilon A_n^{(\delta)}}P(y)P(R_n(X) > 2^{n(H+\epsilon)}|X_1^n = y)\\& + \sum_{y \epsilon {A_n^{(\delta)}}^c}P(y)P(R_n(X) > 2^{n(H+\epsilon)}|X_1^n = y)\\
&< \sum_{y\epsilon A_n^{(\delta)}}P(y)[\xi_y e^{-2^{n(H+\epsilon)}\xi_yP(y)}[1+ K(y,2^{n(H+\epsilon)})\\&+ 2C_y(\xi_y2^{n(H+\epsilon)}P(y)\vee 1)]] + \sum_{y \epsilon {A_n^{(\delta)}}^c}P(y)~~~~~ (a)
\end{split}
\end{equation*}
\begin{equation}\label{main1}
\begin{split}
&< \sum_{y \epsilon {A_n}^{(\delta)}}P(y)[E_2 e^{-2^{n(H+\epsilon)}E_1P(y)}[1+  V\\&+  2d(E_22^{n(H+\epsilon)}P(y)\vee 1)]  +  \sum_{y\epsilon {A_n^{(\delta)}}^c}P(y)~~~~~~~~ (b)
\end{split}
\end{equation}
where
\begin{equation*}
V = 2\sqrt{d(2^{n(H+\epsilon)}E_2P(y)\vee 1)(1+d(2^{n(H+\epsilon)}E_2P(y)\vee 1))}.
\end{equation*}
$(a)$ follows from the use of inequality (\ref{UB}) and Remark 6 as stated in Appendix. $(b)$ follows from using the fact that $\xi_y ~\epsilon~[E_1,E_2]$ and $C_y \rightarrow 0 ~as~n \rightarrow \infty \Rightarrow C_y < d ~ \forall$ $y$ and $n$ large enough, where $d > 0$ is an arbitrary constant. For $y ~\epsilon~ A_n^{(\delta)}$, we have $$2^{n(\epsilon-\delta)}\leq 2^{n(H+\epsilon)}P(y) \leq 2^{n(\epsilon + \delta)}.$$ For every $\epsilon > 0$, choose $\delta = \frac{\epsilon}{2}$. Consequently, we have
\begin{equation}\label{sub1}
2^{\frac{n\epsilon}{2}} \leq 2^{n(H+\epsilon)}P(y) \leq 2^{\frac{3n\epsilon}{2}} ~ \forall y ~\epsilon~A_n^{(\frac{\epsilon}{2})}.
\end{equation}
Also, $2^{\frac{3n\epsilon}{2}}E_2 > 1$ since $E_2 > 1$. Hence, using (\ref{main1}) and (\ref{sub1}) we have,
\begin{equation}\label{final1}
\begin{split}
P(\frac{\log R_n(X)}{n} > H+ \epsilon) \leq \sum_{y\epsilon A_n^{(\frac{\epsilon}{2})}}P(y)[E_2e^{-E_12^{\frac{n\epsilon}{2}}}(1+\\2\sqrt{dE_22^{\frac{3n\epsilon}{2}}(1+dE_22^{\frac{3n\epsilon}{2}})}+ 2dE_22^{\frac{3n\epsilon}{2}})] +  \sum_{y \epsilon {A_n^{(\frac{\epsilon}{2})}}^c}P(y).
\end{split}
\end{equation}
Using (\ref{LDP_Sh}) and (\ref{final1}), for processes having exponential rates for entropy and satisfying $\psi$-mixing condition or $\phi$-mixing condition with summable coefficients, we have
\begin{equation}\label{final_r1}
\begin{split}
&P(\frac{\log R_n(X)}{n} > H+ \epsilon) \leq  \Big\{E_2e^{-E_12^{\frac{n\epsilon}{2}}}[1+ 2dE_22^{\frac{3n\epsilon}{2}}\\&+2\sqrt{dE_22^{\frac{3n\epsilon}{2}}(1+dE_22^{\frac{3n\epsilon}{2}})}]\Big\} + r(\frac{\epsilon}{2},n)
< e^{-f(\epsilon)n}.
\end{split}
\end{equation}
This completes the proof of Theorem 3.\\
{\bf Remark 3:} Since the first term on the right hand side of inequality (\ref{final_r1}) stated above rapidly (super exponentially) converges to $0$, $f(\epsilon)$ behaves in a similar manner as $-\frac{\ln r(\frac{\epsilon}{2},n)}{n} = k(\frac{\epsilon}{2})$. (Also see Remark 7 as stated in Appendix) \\
{\bf Proof of Corollary 2:}
From Observation 1, we have $$R_n(x) > 2^{n(H+\epsilon)} \Leftrightarrow L_{2^{n(H+\epsilon)}}(x) < n $$
\begin{equation*}
\Rightarrow P(L_{2^{n(H+\epsilon)}}(X) < n) = P(R_n(X) > 2^{n(H+\epsilon)}) < e^{-f(\epsilon)n}.
\end{equation*}
$\forall~n\geq N(\epsilon)$.
Now, letting $m = 2^{n(H+\epsilon)}$, we have
\begin{equation*}
\begin{split}
P(L_{m}(X) < \frac{\log m}{H + \epsilon}) < e^{-f(\epsilon)\frac{\log m}{H + \epsilon}}   ~~~~\forall ~m \geq M(\epsilon)\\
\Rightarrow P(\frac{\log m}{L_m(X)} > H+ \epsilon) < e^{-f(\epsilon)\frac{\log m}{H + \epsilon}}.
\end{split}
\end{equation*}
{\bf Proof of Theorem 4:}
Let $A_n^{(\frac{\epsilon}{2})}$ be the same set as considered in the proof of Theorem 3. For each $y~\epsilon~A_n^{(\frac{\epsilon}{2})}$, we have
\begin{equation}\label{sub2}
2^{-\frac{3n\epsilon}{2}} \leq P(y)2^{n(H-\epsilon)} \leq 2^{-\frac{n\epsilon}{2}}.
\end{equation}
Now,
\begin{equation}\label{final2}
\begin{split}
&P(\frac{\log R_n(X)}{n} < H-\epsilon)
 = 1 - P(\frac{\log R_n(X)}{n} \geq H-\epsilon)\\
                                      &\leq 1-P(\frac{\log R_n(X)}{n} > H-\epsilon)\\
                                      &=1-P(R_n(X) > 2^{n(H-\epsilon)})\\
                                      &=1-\sum_{y \epsilon \mathcal{A}^n}P(y) P(R_n(X) > 2^{n(H-\epsilon)}|X_1^n = y)\\
                                      &< 1-\sum_{y\epsilon\mathcal{A}^n}P(y)[\xi_ye^{-P(y)\xi_y2^{n(H-\epsilon)}}(1- \\& 2\sqrt{C_y(\xi_yP(y)2^{n(H-\epsilon)}\vee 1)})]~~(a) \\
                                      &< 1-\sum_{y\epsilon A_n^{(\frac{\epsilon}{2})}}P(y)[\xi_ye^{-E_22^{-\frac{n\epsilon}{2}}}(1- 2\sqrt{C_y(E_22^{-\frac{n\epsilon}{2}}\vee 1)})] ~(b) \\
                                      &= 1- \sum_{y\epsilon A_n^{(\frac{\epsilon}{2})}}P(y)[\xi_ye^{-E_22^{-\frac{n\epsilon}{2}}}(1-2\sqrt{C_y})]~(c)
\end{split}
\end{equation}
Here, $(a)$ follows from (\ref{LB}), $(b)$ follows from the fact that $\xi_y ~\epsilon~[E_1,E_2]$ and inequality (\ref{sub2}). Also in $(b)$ the negative term contributed by sequences belonging to the set ${A_n^{(\frac{\epsilon}{2})}}^c$ is ignored because we are looking at an upper bound. $(c)$ follows because eventually $E_22^{-\frac{n\epsilon}{2}} < 1$, since $E_22^{-\frac{n\epsilon}{2}} \rightarrow 0~(as ~ n~\rightarrow ~\infty).$

To proceed further, we introduce the following notations, let $A_1 = \mathcal{A}^n \backslash B_n(s); A_2 = A_n^{(\frac{\epsilon}{2})}.$ From (\ref{4}) and (\ref{LDP_Sh}), we have $P(A_1) > 1- D_1e^{-d_1n}$ and $P(A_2) > 1- e^{-k(\frac{\epsilon}{2})n}$ respectively for processes with exponential rates for entropy. Let $A_3 = A_1 \cap A_2$.

Therefore from (\ref{final2}), we have
\begin{equation}\label{final_r2_sub}
\begin{split}
&P(\frac{\log R_n(X)}{n} < H -\epsilon) < 1-\\&\sum_{y\epsilon A_3}P(y)\xi_ye^{-E_22^{-\frac{n\epsilon}{2}}}(1-2\sqrt{C_y})-\\&\sum_{y\epsilon A_2\backslash A_3}P(y)\xi_ye^{-E_22^{-\frac{n\epsilon}{2}}}(1-2\sqrt{C_y})\\
&\leq 1-\sum_{y\epsilon A_3}P(y)(1-D_2e^{d_2n})e^{-E_22^{-\frac{n\epsilon}{2}}}(1-2\sqrt{D_o}e^{-\frac{\Gamma n}{2}})~(d)\\
&=1-P(A_3)(1-D_2e^{-d_2n})e^{-E_22^{-\frac{n\epsilon}{2}}}(1-2\sqrt{D_o}e^{-\frac{\Gamma n}{2}}) \\
&\leq 1-\Big[e^{-E_22^{-\frac{n\epsilon}{2}}}(1- (D_1+1)e^{-\min\{d_1,k(\epsilon)\}n})\\&(1-D_2e^{-d_2n})(1-2\sqrt{D_o}e^{-\frac{\Gamma n}{2}})\Big] ~(e)\\
&\leq 1-(1-C'e^{-u(\epsilon)n})e^{-E_22^{-\frac{n\epsilon}{2}}}
\end{split}
\end{equation}
Here, $(d)$ follows from (\ref{xi}) and (\ref{exp_phi}) and ignoring the negative contribution made by the sequences in the set $A_2\backslash A_3$. $(e)$ follows from Lemma 1. $C' > 0$ (constant) and $u(\epsilon)$ (positive valued function $\forall~ \epsilon > 0$ and 0 if $\epsilon = 0$) are obtained after simplification of $(e)$. Now, using extended mean value theorem for the function $e^{-z}$, $$e^{-E_22^{-\frac{n\epsilon}{2}}} = 1-E_22^{-\frac{n\epsilon}{2}}+ \frac{e^{-c}}{2}E_2^22^{-n\epsilon}.$$ Here, $c~\epsilon~(0,E_22^{-\frac{n\epsilon}{2}})$. Therefore, we have
\begin{equation}\label{sub_2}
e^{-E_22^{-\frac{n\epsilon}{2}}} \geq 1-E_22^{-\frac{n\epsilon}{2}}.
\end{equation}
Hence, using (\ref{sub_2}) in (\ref{final_r2_sub}), we get
\begin{equation}\label{final_r2}
P(\frac{\log R_n(X)}{n} < H -\epsilon) < e^{-g(\epsilon)n}.
\end{equation}
where $g(\epsilon)$ is a positive valued function $\forall ~\epsilon~>~0$ and $g(0) = 0$.
This completes the proof of Theorem 4.\\
{\bf Proof of Corollary 3:}
Using Observation 1, we have $$L_{2^{n(H-\epsilon)}}(x) > n \Rightarrow R_n(x) \leq 2^{n(H-\epsilon)}$$
\begin{equation*}
\Rightarrow P(L_{2^{n(H-\epsilon)}}(X) > n)\leq P(R_n(X) \leq 2^{n(H-\epsilon)}) < e^{-g(\epsilon)n}.
\end{equation*}
$\forall ~n \geq N'(\epsilon)$. Now, letting $m = 2^{n(H-\epsilon)}$, we have
\begin{equation*}
\begin{split}
P(L_{m}(X) > \frac{\log m}{H - \epsilon}) < e^{-g(\epsilon)\frac{\log m}{H - \epsilon}}   ~~~~\forall ~m \geq M'(\epsilon)\\
\Rightarrow P(\frac{\log m}{L_m(X)} < H- \epsilon) < e^{-g(\epsilon)\frac{\log m}{H - \epsilon}}.
\end{split}
\end{equation*}
{\bf Remark 4:} Note that in the first step in Eq. (\ref{final2}) we have a term $1-  P(\frac{\log R_n(X)}{n} > H-\epsilon) = P(R_n(X) \leq 2^{n(H-\epsilon)})$. Further, in the proof of Theorem 4, the bound $e^{-g(\epsilon)n}$ is obtained on this term. Hence, there is no ambiguity in using the exponential bound obtained in Theorem 4 on $P(R_n(X) \leq 2^{n(H-\epsilon)})$.
\section{Estimator for Entropy}
Motivated by experimental results on estimators based on match lengths given in \cite{Suhov98}, we propose an estimator based on \emph{recurrence times} as given below:\\
{\bf  Estimator:}
Consider $R_{n,i}(X) = R_n(T^iX).$ \\Define: $ J_n(X) = \frac{1}{Q(n)}\sum_{i = 1}^{Q(n)} \frac{\log R_{n,i}(X)}{n}$.\\
{\bf Proposition 1:} If $Q(n)$ is of the polynomial order, then for processes which are exponentially $\phi$-mixing and have exponential rates for entropy,  $\lim_{n \rightarrow \infty} J_n(X) = H  ~a.s. $
with $J_n(X)$ satisfying large deviation property. The proof of the proposition is given below:
\begin{equation}\label{RHS_Estimator}
\begin{split}
P(J_n(X) > H+ \epsilon ) &= P(\frac{1}{Q(n)}\sum_{i=1}^{Q(n)}\frac{\log R_{n,i}(X)}{n} > H + \epsilon)\\
                                        &\leq \sum_{i=1}^{Q(n)}P(\frac{\log R_{n,i}(X)}{n} > H + \epsilon)~(a)\\
                                        &< \sum_{i=1}^{Q(n)}e^{-f(\epsilon)n}   ~~~~\forall~ n \geq  N(\epsilon) ~(b)\\
                                        &= Q(n)e^{-f(\epsilon)n}.
\end{split}
\end{equation}
Here, step $(a)$ follows from Remark 8 given in appendix and step $(b)$ follows from the stationarity of the source $X$ and Theorem 3.
Similarly,
\begin{equation}\label{LHS_Estimator}
\begin{split}
P(J_n(X) < H- \epsilon )  &= P(\frac{1}{Q(n)}\sum_{i=1}^{Q(n)}\frac{\log R_{n,i}(X)}{n} < H - \epsilon)\\
                                        &\leq\sum_{i=1}^{Q(n)}P(\frac{\log R_{n,i}(X)}{n} < H - \epsilon) ~(a)\\
                                        &< \sum_{i=1}^{Q(n)}e^{-g(\epsilon)n}   ~~~~\forall~ n \geq N'(\epsilon)  ~ (b)\\
                                        &= Q(n)e^{-g(\epsilon)n}.
\end{split}
\end{equation}
Here, step $(a)$ follows from Remark 8 and step $(b)$ follows from the stationarity of the source $X$ and Theorem 4.
Therefore, combining (\ref{RHS_Estimator}) and (\ref{LHS_Estimator}), we have
\begin{equation}\label{Tail_Estimator}
P(|J_n(X)- H| > \epsilon ) < 2Q(n)e^{-I(\epsilon)n} ~\forall~ n \geq N''(\epsilon)
\end{equation}
where $N''(\epsilon) = \max\{N(\epsilon), N'(\epsilon)\}$.\\
For $Q(n)$ of polynomial order, we have
\begin{equation}\label{BC_Estimator}
\begin{split}
\sum_{n = 1}^{\infty}P(|J_n(X)- H| > \epsilon ) &< \sum_{n=1}^{N''(\epsilon)-1}P(|J_n(X)-H| >\epsilon)\\& +\sum_{n=N''(\epsilon)}^{\infty}2Q(n)e^{-I(\epsilon)n}\\ &< N''(\epsilon) + \sum_{n=N''(\epsilon)}^{\infty}2Q(n)e^{-I(\epsilon)n}\\                                                                         &<\infty.
\end{split}
\end{equation}
Hence, by Borel-Cantelli Lemma
\begin{equation}
\lim_{n\rightarrow \infty} J_n(X) = H ~~~~~~~ a.s.
\end{equation}
{\bf Remark 5:} The bounds we establish on convergence rates are loose, we \emph{conjecture} that our proposed estimator will converge to entropy rate at a faster rate than $2e^{-I(\epsilon)n}$.
\section{Conclusion}
In this paper, we have proved the Large deviation property for the normalized version of recurrence times for exponentially $\phi$-mixing processes. Further, we have also shown this property to hold for our proposed estimator of entropy based on recurrence times. As a future work, it will be interesting to answer if there are faster rate functions than $f(\epsilon)$ and $g(\epsilon)$ in this context, and further on what more classes of processes large deviation property holds for normalized version of recurrence times. Also, one can conduct experimental or theoretical studies comparing the convergence rates of the estimator based on match length given in \cite{Suhov98} and that based on recurrence times proposed in this paper.
\section*{Appendix}
{\bf Remark 6:} Note that in step (a), inequality (\ref{UB}) has been used, however it is important to check if it can be applied. This is verified below:  $$\rho_y = \frac{2\sqrt{C_y}}{(\sqrt{1+C_y}+\sqrt{C_y})\xi_yP(y)}~~\forall~ y ~\epsilon~ A_n^{(\delta)}$$
Using lower bounds on $P(y)$ and $\xi_y$, we have $$\rho_y \leq \frac{2\sqrt{C_y}~2^{n(H+\delta)}}{(\sqrt{1+C_y}+\sqrt{C_y})E_1}~~\forall~y~\epsilon~A_n^{(\delta)}.$$
Since $C_y \rightarrow 0~ (as ~n\rightarrow\infty)$, for a given $d' > 0$, $\frac{2\sqrt{C_y}}{\sqrt{1+C_y}+\sqrt{C_y}} < d'$ for $n$ large enough. Now, we choose $d'$ such that $ 0<d' < E_1$. Since eventually $\delta$ is chosen to be less than $\epsilon$, we have $$ \rho_y < 2^{n(H+\delta)}<2^{n(H+\epsilon)} ~\forall~ y ~\epsilon ~A_n^{(\delta)}.$$
{\bf Remark 7:} Note that, though we prove Theorem 3 under the restriction of certain mixing conditions and using inequality (\ref{UB}), it can also be proved using Markov Inequality and Kac's Lemma under no restriction of mixing. However, the super exponential behavior shown by first term in the proof of Theorem 3 (see Inequality (\ref{final_r1})) is not evident from this alternative proof for mixing sources considered. Due to space limitations, we have omitted this proof.\\
{\bf Remark 8:} Let $Z_1, Z_2,....,Z_m$ be $m$ real valued random variables. Consider the following probability, $P(\frac{1}{m}\sum_{i=1}^m Z_i  > r)$ and set $E_i = \{\omega: Z_i(\omega) > r\}$. Now,
\begin{equation*}
\begin{split}
P(\frac{1}{m}\sum_{i=1}^m Z_i  > r) &\leq P(\cup_{i=1}^m E_i)\\
                                                             &\leq \sum_{i=1}^m P(E_i)~~(Union~Bound) \\
                                                             &=\sum_{i=1}^mP(Z_i > r).
\end{split}
\end{equation*}
Similarly, by changing `$>$' sign with `$<$' accordingly, it can be proved that
$$P(\frac{1}{m}\sum_{i=1}^m Z_i < r) \leq \sum_{i=1}^mP(Z_i < r).$$
{\bf Proof of Lemma 1:}
\begin{equation*}\label{Lemma1}
\begin{split}
P(A_1\cup A_2)  &\leq 1 \Rightarrow P(A_1)+P(A_2)-P(A_1\cap A_2) \leq 1\\
                   &\Rightarrow P(A_1\cap A_2) \geq P(A_1) + P(A_2) - 1\\
                   &\Rightarrow P(A_1\cap A_2) > 1-p_1e^{-p_2n}+ 1-q_1e^{-q_2n} - 1\\
                   &\Rightarrow P(A_1\cap A_2) > 1-(p_1e^{-p_2n}+q_1e^{-q_2n})\\
                   &\Rightarrow P(A_1\cap A_2) > 1- (p_1+q_1)e^{-\min\{p_2,q_2\}n}.
\end{split}
\end{equation*}

\end{document}